# Using ADP to Understand and Replicate Brain Intelligence: the Next Level Design

Paul J. Werbos[1]
Room 675, National Science Foundation
Arlington, VA 22203, US
pwerbos@nsf.gov

Abstract-Since the 1960's I proposed that we could understand and replicate the highest level of intelligence seen in the brain, by building ever more capable and general systems for adaptive dynamic programming (ADP) – like "reinforcement learning" but based on approximating the Bellman equation and allowing the controller to know its utility function. Growing empirical evidence on the brain supports this approach. Adaptive critic systems now meet tough engineering challenges and provide a kind of first-generation model of the brain. Lewis, Prokhorov and I have done some work on second-generation designs. I now argue that mammal brains possess three core capabilities – creativity/imagination and ways to manage spatial and temporal complexity - even beyond the second generation. This chapter reviews previous progress, and describes new tools and approaches to overcome the spatial complexity gap. The Appendices discuss what we can learn about higher functions of the human mind from this kind of mathematical approach.

## I. INTRODUCTION

No one on earth today can write down a complete set of equations, or software system, capable of learning to perform the complex range of tasks that the mammal brain can learn to perform. From an engineering viewpoint, this chapter will provide an updated roadmap for how to reach that point. From a neuroscience viewpoint, it will provide a series of *qualitative* but *quantifiable* theories of how intelligence works in the mammal brain. The main text of the chapter was written for an engineering workshop; it explains the basic mathematical principles, and how they relate to some of the most important gross features of the brain. A new appendix, written for arxiv.org, discusses more of the implications for comparative neuroscience and for our subjective experience as humans.

The main text of this chapter will not address the human mind as such. In nature, we see a series of *levels* of intelligence or consciousness [1]; for example, within the vertebrates, M. E. Bitterman [2] has shown that there are major qualitative jumps from the fish to the amphibian, from the amphibian to the reptile, and from the reptile to even the simplest mammal. Ninety-nine percent of the higher parts of the human brain consist of structures, like the six-layer neocortex, which exist in the smallest mouse, and show similar general-purpose-learning abilities in the mouse; therefore, the scientific goal of understanding the kind of learning and intelligence that we see in the smallest mouse is an important step towards understanding the human mind, but it is certainly not the whole thing.

Section II of this chapter will briefly review the general concepts of optimization and ADP. It will give a few highlights from the long literature on why these offer a central organizing principle both for understanding the brain and for improving what we can do in engineering. Section III will review the first and second generation ADP designs, and their relevance to brain-style intelligence. Section IV will discuss how to move from second-generation designs to the level of intelligence we see in the brain of the smallest mouse – with a special emphasis on how to handle spatial complexity, by learning symmetry groups, and to incorporate that into an ADP design or larger brain. Appendices A-C go on to discuss *comparative* cognitive neuroscience (other classes of vertebrate brain), and the significance of this work to understanding higher capabilities of the human brain and subjective experience. Appendix D gives a brief explanation of how intelligence at the level of a bird brain could do far better than today's systems for managing electric power grids. Appendix E discusses the neural mechanisms for imagination and creativity, and suggests a new hypothesis for their evolutionary origins.

---

[1] [1]The views herein represent no one's official views, but the chapter was written on US government time

# II. WHY OPTIMALITY AND WHY ADP?

A.  *Optimality As An Organizing Principle for Understanding Brain Intelligence*

For centuries, people have debated whether the idea of optimization can help us understand the human mind. Long ago, Aristotle proposed that all human efforts and thought are ultimately based on the pursuit of (maximizing) happiness – a kind of inborn "telos" or ultimate value. Utilitarians like John Stuart Mill and Jeremy Bentham carried this further. A more updated version of this debate can be found in [3] and in the chapter by Perlovsky; here I will only review a few basic concepts.

To begin with [4], animal behavior is ultimately about choices as depicted here:.

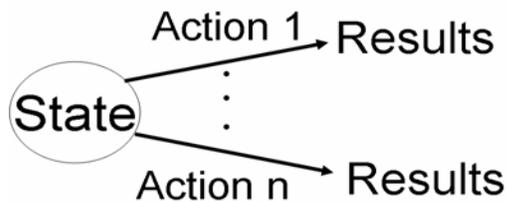

The simplest types of animals may be born with fixed rules about what actions to take, as a function of the state of their environment as they see it. More advanced animals, instead, have an ability to select actions which are somehow based on the *results* that the actions might have. *Functionality* of the brain is about making choices which yield better results. *Intelligence* is about *learning* how to make better choices

To put all this into mathematics, we must have a way to evaluate which results are "better" than which other results. Von Neumann's concept of Cardinal Utility function [5] provides that measure; it is the foundation of decision theory [6], risk analysis, modern investment analysis, and dynamic programming, among others. Before Von Neumann, most economists believed that (consumer) values or preferences are strictly an *ordinal* relationship; for example, you can prove by experiment that most people prefer steak over hamburger, and that they prefer hamburger over nothing, but it seemed meaningless to argue that "the gap between hamburger and nothing is bigger than the gap between steak and hamburger." Von Neumann showed that we *can* make these kinds of numerical comparisons, once we introduce probabilities into our way of thinking. For example, given a 50-50 coin flip between steak or nothing, versus a certainty of hamburger, many people prefer hamburger. Of course, a more complete utility function for human consumers would account for other variables as well.

Usually, when we talk about discrete "goals" or "intentions," we are not talking about the long-term values of the organism. Rather, we are talking about subgoals or tactical values, which are intended to yield better results or outcomes. The utility function which defines what is "better" is the foundation of the system as a whole.

Next consider the analogy to physics.

In 1971-1972, when I proposed a first generation model of intelligence, based on ADP, to a famous neuroscientist, he objected: "The problem with this kind of model is that it creates an anthropomorphic view of how the brain works. I have spent my entire life teaching people how to overcome a bachtriomorphic view of the frog. Thinking about people by using empathy could be a disaster for science. Besides, even in physics, we know that the universe is maximizing a kind of utility function, and we don't think of the universe in anthropomorphic terms."

From a strictly objective viewpoint, his argument actually supports the idea of trying to use optimization as a central organizing principle in neuroscience. After all, if it works in physics, in a highly rigorous and concrete way, why not here? If we can unify our functional understanding of the brain not only with engineering, but with subjective experience and empathy, isn't this a source of strength, so long as we keep track of which is which? In fact, my real goal here has been to develop the kind of mathematical understanding which really does help us to unify our cybernetic understanding, our objective understanding of brain and behavior, and our subjective understanding of ourselves and others.

But does it really work that way in physics? Partly so. According to classical physics, the universe really does solve the optimization problem depicted here:

```
_______________________________  φ(x, t₊)
      φ(x, t)    t₊ > t > t₋
_______________________________  φ(x, t₋)
```

The universe has a kind of "utility function," $\mathscr{L}(\mathbf{x}, t)$. It "chooses" the states φ of all particles and fields at all times t by choosing states which maximize the total sum of $\mathscr{L}$ across all of space time, between time $t_-$ and time $t_+$, subject to the requirement that they provide a continuous path from the fixed state at some initial time $t_-$ and some final time $t_+$. This elegant formalism, due to Lagrange, provides a very simple parsimonious description of the laws of physics; instead of specifying n dynamic laws for n types of particle or field, we can specify the "Lagrangian function $\mathscr{L}$," and derive all the predictions of physics from there. In order to perform that calculation, we can use an equation from classical physics, the Hamilton-Jacobi equation, which tells us how to solve deterministic optimization problems across time or space-time.

But that is not the whole story. Hamilton and Lagrange had many debates about whether the universe really maximizes $\mathscr{L}$ – or does it minimize it or find a minmax solution? Does the physical universe find something that looks like the outcome of a two-person zerosum game? By the time of Einstein, it appeared so. Modern quantum theory gets rid of the deterministic assumption, but adds random disturbance in a very odd way. It actually turns out that we can recover something like Lagrange's original idea, which fits the tested predictions of modern quantum theory, by introducing a stochastic term whose statistics are symmetric both in space and in time; however, the details are beyond the scope of this chapter. (See www.werbos.com/ reality.htm.)

To describe the brain, it is not enough to use the old optimization rule of Hamilton and Jacobi. We need to consider the stochastic case, because animals, like us, cannot predict our environment in a deterministic way. The foundation for optimization over time in the stochastic case is the Bellman equation, a great breakthrough developed by Bellman in 1953, made possible by Von Neumann's concept of Cardinal Utility function.

The principles of optimality are important to fundamental physics – but also to thermo-dynamics, and to the physics of emergent phenomena in general. Those details are beyond the scope of this chapter.

Finally, let me address two of the most common questions which people tend to ask when I talk about the brain as an "optimization machine."

First: If brains are so optimal, why do humans do so many stupid things? Answers: Brains are designed to *learn* approximate optimal policy, as effectively as possible with *bounded computational resources* (networks of neurons), starting from a less optimal start. They never learn to play a perfect game of chess (nor will our computers, nor will any other algorithm that can be implemented on a realistic computer) because of constraints on computational resources. *We just do the best we can*.
Also, when one human (a researcher) criticizes another, we are seeing a comparison between *two* highly intelligent systems. Some brains learn faster than others. In my view, humans themselves are an intermediate state towards an even higher/faster intelligence, as discussed in the Appendices of this paper.

Second question: if this optimization theory of the brain is correct, wouldn't brains get stuck in local minima, just like artificial optimization programs when confronted with a complex, nonlinear environment? Answers: they do indeed. Every person on earth is caught in a "local minimum," or rut, to some degree. In other words, we could all do a bit better if we had more creativity. But look at those hairy guys (chimpanzees) in the jungle, and the rut they are in!

The optimization theory of the brain implies that our brains *combine* an incremental learning ability with an ability to learn to be more creative – to do better and better "stochastic search" of the options available to us. There are a few researchers in evolutionary computing or stochastic search who tell us that their algorithms are guaranteed to find the global optimum, eventually; however, those kinds of guarantees are not very realistic because, for a system of realistic complexity, they require astronomical time to actually get to the optimum.

B.  *Optimality and ADP In Technology*

The benefits of adaptive dynamic programming (ADP) to technology have been discussed by many other authors in the past[7], with specific examples. Here I will review only a few highlights. (See also the example of the electric power grid, discussed at the end of the Appendix.)

Many control engineers ask: "Why try to find the optimal controller out of all possible controllers? It is hard enough just to keep things from blowing up – to stabilize them at a fixed point." In fact – the most truly stable controllers now known are nonlinear feedback controllers, based on "solving" the "Hamilton-Jacobi-Bellman" equation. But in order to implement that kind of control, we need mechanisms to "numerically solve" (approximate) the Bellman equation as accurately as possible. ADP is the machinery to do that.

Furthermore – there are times when it is impossible to give a truly honest absolute guarantee of stability, under accurate assumptions. Certainly, a mouse running through the field has no way to guarantee its survival – nor does the human species as a whole, in the face of the challenges now confronting us. (See www.werbos.com.) In that kind of real-world situation, the challenge is to *maximize the probability* of survival; that, in turn, is a stochastic optimization problem, suitable for ADP, and not for deterministic methods. (Recent work by Gosavi has explored that family of ADP applications.) Verification and validation for real complex systems in the real world is heavily based on empirical tests and statistics already.

Finally, in order to address nonlinear optimization problems in the general case, we absolutely must use universal nonlinear function approximators. Those could be Taylor series – but Barron showed years ago that the simplest form of neural networks offer more accurate nonlinear approximation that Taylor series or other linear basis function approximators, in the general case, when there is a large number of state variables. Use of more powerful and accurate approximators (compatible with distributed hardware, like emerging multicore chips) is essential to more accurate approximations and better results.

## III. FIRST AND SECOND GENERATION ADP MODELS OF BRAIN INTELLIGENCE

A.  *Origins and Basics of the First Generation Model*

Backpropagation and the first true ADP design both originated in my work in the 1970's, as shown in the following simplified flow chart:

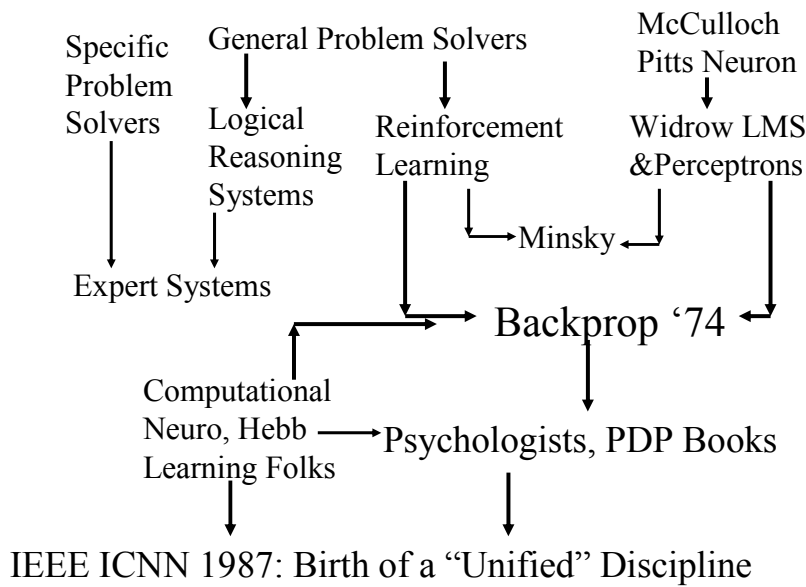

# Where Did ANNs Come From?

In essence, the founders of artificial intelligence (AI) – Newell, Shaw and Simon, and Minsky [8] – proposed that we could build brain-like intelligent systems by building powerful reinforcement learning systems. However, after a great deal of experimentation and intuition and heuristic thinking, they could not design systems which could optimize more than a few variables. Knowing that the brain can handle many thousands of variables, they simply gave up – just as they gave up on training simplified neural models (multilayer perceptrons). Amari, at about the same time, wrote that perhaps derivatives might be used somehow to train multilayer perceptrons – but suggested that it would be unlikely to work, and did not provide any algorithm for actually calculating the required derivatives in a distributed, local manner.

In 1964, I – like many others – was deeply inspired by Hebb's classic book on intelligence [9]. Inspired by the empirical work on mass action and learning in the brain (by Lashley, Freeman, Pribram and others), he proposed that we would not really need a highly complex model in order to explain or reproduce brain-like intelligence. Perhaps we could generate intelligence as the emergent result of learning; we could simply construct billions of model neurons, each following a kind of universal neuron learning rule, and then intelligence could emerge strictly as a result of learning. I tried very hard to make that work in the 1960's, and failed. The key problem was that Hebb's approach to a universal learning rule is essentially calculating correlation coefficients; those are good enough to construct useful associative memories, as Grossberg showed, but not to make good statistical predictions or optimal control. They are simply not enough by themselves to allow construction of an effective general-purpose reinforcement learning machine.

By 1971-1972, I realized that Hebb's vision could be achieved, if we relax it only very slightly. It is possible to design a general purpose reinforcement learning machine, if we allow just three types of neuron and three general neuron learning rules, instead of just one.

Actually, the key insight here came in 1967. In 1967 (in a paper published in 1968 [4]), I proposed that we could overcome the problems with reinforcement learning by going back to basic mathematical principles – by building systems which learn to approximate the Bellman equation. Use of the Bellman equation is still the only exact and efficient method to compute an optimal strategy or policy of action, for a general nonlinear decision problem over time, subject to noise. The equation is:

$$J(\underline{x}(t)) = \max_{\underline{u}(t)} <U(\underline{x}(t), \underline{u}(t)) + J(\underline{x}(t+1))>/(1+r),$$

where $\underline{x}(t)$ is the state of the environment at time t, $\underline{u}(t)$ is the choice of actions, U is the cardinal utility function, r is the interest or discount rate (exactly as defined by economics and by Von Neumann), where the angle brackets denote expectation value, and where J is the function we must solve for in order to derive the optimal strategy of action. In any state $\underline{x}$, the optimal $\underline{u}$ is the one which solves the optimization problem in this equation. A learning system can learn to approximate this policy by using a neural network (or other universal approximator) to approximate the J function and other key parts of the Bellman equation, as shown in the next figure, from my 1971-1972 thesis proposal to Harvard:

## 1971-2: Emergent Intelligence Is Possible If We Allow Three Types of Neuron (Thesis, Roots)

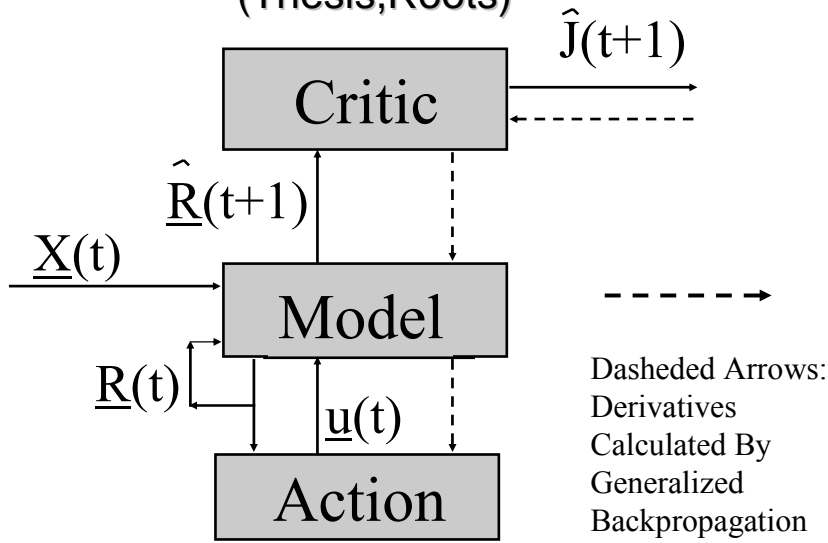

Dasheded Arrows: Derivatives Calculated By Generalized Backpropagation

In that design, I needed a *generalized* form of backpropagation as a tool to calculate the essential derivatives or sensitivity coefficients needed to allow correct incremental learning of all three parts. I formulated and proved a new chain rule for "ordered derivatives" which makes it possible to compute the required derivatives exactly through any kind of large nonlinear system, not just neural networks.

Intuitively, the "Action" network here is the one which actually computes or decides the actions to be taken by the organism. The "Model" network learns how to predict changes in the environment of the organism, and it also learns how to estimate the objective state of reality (**R**), which is more than just the current sensory input (**X**) to the organism. The "Critic" network estimates the "J" function, which is a kind of *learned* value function, similar to what is called "secondary reinforcement" in animal learning.

For my PhD thesis (reprinted in entirety in [10]), I included the proof, and many applications of backpropagation to systems other than neural networks. In [11,12], I described how *generalized backpropagation* can be used in a wide variety of applications, including ADP with components that could be neural networks or *any other* nonlinear differentiable system.

The method which I proposed to adapt the Critic network in 1971-1972 I called "Heuristic Dynamic Programming" (HDP). It is essentially the same as what was later called "the Temporal Difference Method." But I learned very early that the method does not scale very well, when applied to systems of even moderate complexity. It learns too slowly. To solve this problem, I developed the core ideas of two new methods – dual heuristic programming (DHP) and Globalized DHP (GDHP) – published in a series of papers from 1977 to 1981 [13-15]. To prove convergence, in [12], I made small but important changes in DHP; thus [12] is the definitive source for DHP proper. For more robust extensions, see the final sections of [16]. See [7] and [12] for reviews of practical applications of HDP, DHP, GDHP and related adaptive critic systems.

From the figure, you can see that I assumed *discrete* time – a "clock" – in the first generation model of the brain. For decades, it has been known that the cerebral cortex is "modulated" (clocked) by regular timing signals from sources outside the cortex, such as the nonspecific thalamus. It has been known that the output of many types of neurons takes the form of "bursts" at regular time intervals (commonly 100 milliseconds), of continuously varying intensity – not 1's and 0's! Recently it has been discovered that pacemaker cells exist *within* the cerebral cortex, which appear to provide a kind of higher-frequency clock, allowing calculations at discrete times in a kind of "inner loop." Some brain calculations, such as associative memory operations, may be even faster, and may be performed by asynchronous circuits or even molecular mechanisms within cells; however, the "clocks" are essential to the higher-level integrated systems which make use of such subsystems. As I prepare this paper for submission, I have been informed of a major project to build circuits of living neurons on a chip, which may have failed because of failure to include pacemaker cells in the culture or because they looked for a "spiking" code instead of "bursting' code.

*B. A Second Generation Model/Design for Brain-Style Intelligence*

By 1987, I realized that the brain has certain capabilities beyond what any of these first-generation designs offer. Thus I proposed [17] a second generation theory. Key details of that theory were worked out in [12] and in papers with Pellionisz, See the flow chart below and the papers posted at www.werbos.com.

## 2nd Generation "Two Brains in One Model"

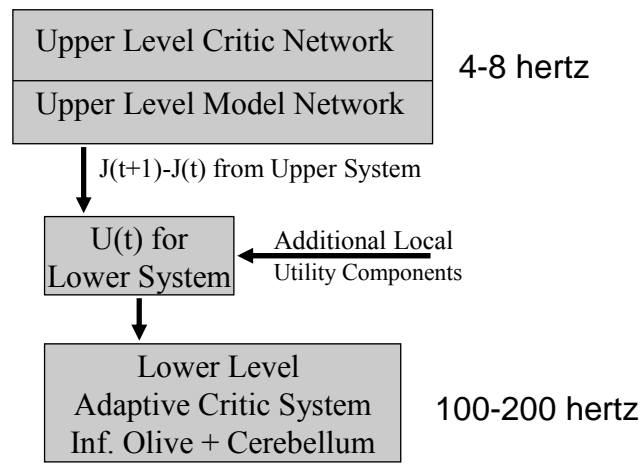

The second generation design was motivated in part by trying to understand the brain. It was also motivated by an engineering dilemma. The dilemma is that truly powerful foresight, in an ADP system, requires the use of Critic Networks and Model Networks which are far more powerful than feedforward neural networks (or Hebbian networks or Taylor series or linear basis function networks). It requires the use of recurrent networks, including networks which "settle down" over many cycles of an inner loop calculation before emitting a calculation. That, in turn, requires a relatively low sampling rate for calculation; about 4-8 frames per second is the rate observed for the cerebral cortex of the mammal brain, in responding to new inputs from the thalamus, the "movie screen" watched by the upper brain. However, smooth muscle control requires a much higher bandwidth of control; to achieve that, I proposed that the brain is actually a kind of master-slave system. In chapter 13 of [12], I provided equations for an "Error Critic for motor control" which provide one possible design for a fast model-free "slave" neural network, matching this model. In that design, the "memory neurons" which estimated the vector **R** are in the Action Network; I proposed that these are simply the Purkinje cells of the cerebellum. They are trained by a kind of distributed DHP-like Critic system, based partly in the inferior olive and partly near the Purkinje cells themselves.

Danil Prokhorov, in various IJCNN papers, showed how that kind of fast design (and some variations he developed) works well, by certain measures, in computational tests. Recent formal work in Frank Lewis's group at the University of Texas (ARRI) has shown strong stability results for continuous-time model free ADP designs which *require* an external value input, exactly like what this master-slave arrangement would provide.

Intuitively… the "master" is like the coach within you, and the "slave" is like the inner football player. The football player has very fast reflexes, and is essential to the game, but he needs to strive to go where the more far-seeing coach sends him. The coach can learn more complex stuff faster than the football player, and responds to a more complex strategic picture. Lower-level stability is mainly provided by the football player.

In 1987, Richard Sutton read [17], and arranged for us to discuss it at great length in Massachusetts. This was the event which injected the idea of ADP into the reinforcement learning school of AI. The paper is cited in Sutton's chapter in [18], which includes an implementation of the idea of "dreaming as simulation" discussed in [17].

*C. Engineering Roadmap and Neuroscience Evidence for Second Generation Theory/Design*

In 1992, I believed that we could probably replicate the level of intelligence we see in the basic mammal brain, simply by refining and filling in these first and second generation theories of how the brain works. In fact, the first and second generation design already offer potential new general-purpose adaptive capabilities far beyond what we now have in engineering. It is still essential that we continue the program of refining and understanding and improving these classes of designs as far as we can go – both for the sake of engineering, and as a prerequisite to set the stage for even more powerful designs.

I have suggested that half of the funding aimed at reverse engineering the brain should still go towards the first and second generation program – half towards the ADP aspects, and half towards the critical subsystems for prediction, memory and so on. (See www.eas.asu.edu/~nsfadp .) Because those are complicated issues, and I have written about them elsewhere, I will not elaborate here.

More and more evidence has accumulated suggesting that optimization (with a predictive or "Model" component) is the right way to understand the brain. For example, Nicolelis and Chapin, in *Science*, reported that certain cells in the thalamus act as advance predictors of other cells. More important, when they cut the existing connections, the thalamo-cortical system would adapt in exactly the right way to relearn how to predict. This is clear evidence that the thalamo-cortical system – the biggest part of the brain – is in great part an adaptive "Model" network, a general-purpose system for doing adaptive "system identification" (as we say in control theory). Barry Richmond has observed windows of forwards and backwards waves of information in this circuit, fully consistent with our Time-Lagged Recurrent Network (TLRN) model of how such a Model network can be constructed and adapted.

Papez and James Olds senior showed decades ago how cells in the "limbic system" convey "secondary reinforcement signals," exactly as we would predict for an adaptive Critic component of the brain. More recent work on the dopamine system in the basal ganglia suggests even more detailed relations between reinforcement learning and actual learning in neural circuits.

A key prediction of the engineering approach has always been the existence of subcircuits to compute the derivatives – the generalized backpropagation – required here. When we first predicted backwards synapses, to make this possible, many ridiculed the engineering approach. But later, in Science, Bliss et al reported a "mysterious" but strong reverse NMDA synapse flow. Spruston and collaborators have reported backpropagation flows (totally consistent with the mathematics of generalized backpropagation) in cell membranes. The synchronized clock signals implied in these designs are also well-known at present to "wet," empirical neuroscientists.

More details – and the empirical implications which cry out for follow-on work – are discussed in some of the papers on my web page, such as papers for books edited by Pribram.

One interesting recent thought: From engineering work, we have learned that the complexity of the *learning* system needed to train a simple input-output system or learned policy is far greater than the complexity of the input-output system itself. A simple example comes from Kalman filtering, where the "scaffolding" matrices (P, etc.) needed for consistent filtering are n times as large as the actual state estimates themselves; n is the number of state variables. Another example comes from the neural network field, where Feldkamp and Prokhorov have shown that Time-Lagged Recurrent Networks (TLRN) do an excellent job of predicting engine behavior – even across different kind of engines. It performs far better than the direct use of Extended Kalman Filters (EKF) in state estimation, and performs as well as full particle filter methods, which are far more expensive (especially when there are many variables in the system, as in the brain). To train these TLRNS, they use backpropagation through time *combined with* "Distributed EKF (DEKF) training," a kind of training which requires updating matrices. Could it be that "junk DNA" includes a large system whose purpose is to tune the adaptation of the "coding DNA," which are after all only a small portion of our genetic system? Could it be that individual neurons do contain very complex molecular memories after all – memories invisible to our conscious mind, but essential to more efficient learning (such as the matrices for DEKF learning)? These are important empirical issues to explore.

## IV. BRIDGING THE GAP TO THE MAMMAL-BRAIN LEVEL

AI researchers like Albus [19] have long assumed that brains must have very complex, explicit, hard-wired hierarchies of systems to handle a high degree of complexity in space and in time. By 1997, I became convinced that they are partly right, because I was able to formulate modified Bellman equations which allow much faster learning in cases where a state space can be sensibly partitioned in a (learnable) hierarchical way[20,21]. Nature would not neglect such an opportunity – and it fit well with emerging new knowledge about the basal ganglia, and ideas from Pribram.

Recent biological data does not support the older hierarchy ideas form AI, but it clearly call out for some kind of specific mechanisms in three core areas: (1) a "creativity/imagination" mechanism, to address the nonconvex nature of complicated optimization problems; (2) a mechanism to exploit modified Bellman equations, in order to cope with the complexity of decisions across multiple time scales; and (3) a mechanism to handle spatial complexity. (I will discuss them here in that order; however, in implementing these capabilities, it is much easier to go in reverse order, starting with spatial complexity.)

3rd Generation View of Creativity/Imagination: Layer V = "Option Networks"

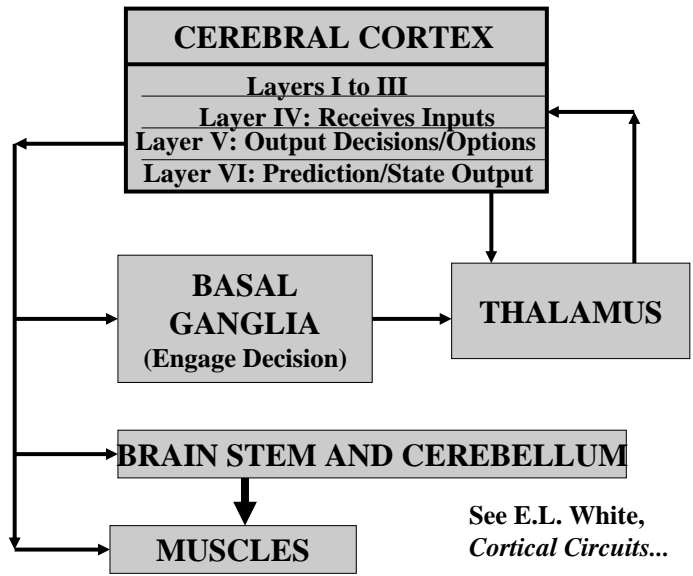

- Challenge: www.werbos.com/WerbosCEC99.htm.
- Important work by Serpen, Pelikan, Wunsch, Thaler, Fu – but still wide open. Widrow testbed.

The flow chart above summarizes a strawman model of the creativity mechanism which I proposed in 1997 [20]. I hoped to stimulate broad research into "brain-like stochastic search." (See my web page for a CEC plenary talk on that challenge). Wunsch, Serpen, Thaler, Pelikan and Fu's group at Maryland have all done important early work relevant to this task, but it hasn't really come together. Likewise, work in the last few years on temporal complexity has not done full justice to the modified Bellman equations, and has not shown as much progress as hoped for; part of the problem is that temporal complexity is usually associated with spatial complexity as well. Also, there is new evidence from neuroscience which has not yet been assimilated on the technology side. (See Appendices B and E.)

The most exciting opportunity before us now is to follow up on more substantial progress and new ideas related to spatial complexity.

An important early clue towards spatial complexity came from the work of Guyon, LeCun, and others at AT&T, illustrated below:

# Moving Window Net: Clue Re Complexity

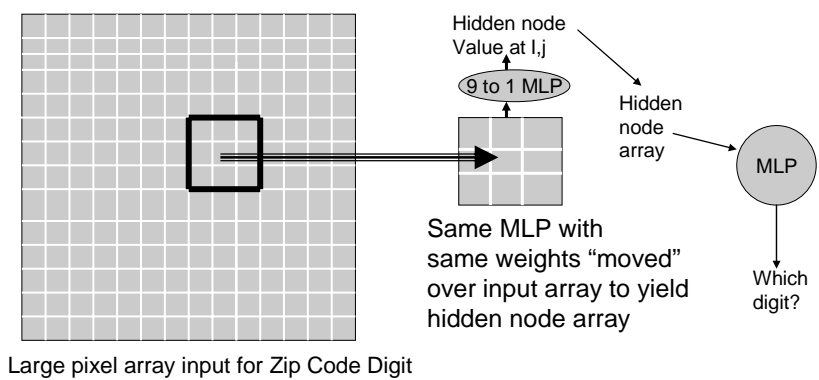

Large pixel array input for Zip Code Digit

- Best ZIP Code Digit Recognizer Used "Moving Window" or "conformal" MLP! (Guyon, LeCun, AT&T story, earlier…)
- Exploiting symmetry of Euclidean translation crucial to reducing number of weights, making large input array learnable, outcomes.

The most accurate ZIP code digit recognizer then came from a simple MLP network, *modified* to exploit symmetry with respect to spatial translation. Instead of independently training hidden neurons to process pieces of an image, they would train a *single* hidden neuron, and *re-use* it in different locations, by moving it around the image. LeCun later called this a "conformal neural network." He had excellent results training it by backpropagation in many image processing tasks. Nevertheless, these feed-forward networks could still not learn the more complex kinds of mappings, like the connectedness mapping described long ago by Minsky [22]; it is not surprising that a network which could not handle connectedness, or learn to emulate Hough relaxation of image data, could not learn how to segment an entire ZIP code.

In 1994, Pang and I demonstrated a network that could solve these problems – a "Cellular SRN," (CSRN), which combines the key capabilities of a Simultaneous Recurrent Network [12] and a "conformal" network. This immediately allows prediction and control and navigation through complex two-dimensional scenes (and images) far more complex than an MLP could ever learn. That did not become immediately popular, in part because the learning was slow and tools were not available to make it easy for people to take advantage of the great brain-like power of such networks. This year, however, Ilin, Kozma and I reported [23] a new learning tool which dramatically speeds up learning, and is available from Kozma as a MatLab tool. This by itself opens the door for neural networks to solve complex problems they could never really handle in the past. (There is great room for research to speed it up even more, but it is ready for practical use already.)

In 1997 [20] and in subsequent tutorials (and a patent), I proposed a more general approach to exploiting symmetry, which I called the ObjectNet. Instead of mapping a complex input field into M rectangular cells, all governed by a common "inner loop" neural network, one may map it into a network of k types of "Objects," with k different types of "inner loop" neural networks. This has great potential in areas like electric power and image processing, for example. A conventional MLP or recurrent network can learn to manage perhaps a few dozen variables in a highly nonlinear system – but how can one design a neural network which inputs the thousands of variables of an entire electric power grid and predict the system as a whole? Object nets provide a way of doing that. The ObjectNet is essentially a functional, explicit way to implement the classic notion from cognitive science of a "cognitive map."

This year, Venayagamoorthy has published preliminary results showing how an ADP system based on a simple, feedforward version of ObjectNet can handle power systems more complex than the earlier first-generation brain-like systems (which already outperformed more conventional control methods). More astonishing – David Fogel used a simple ObjectNet as the Critic in a system adapted to play chess. This was the world's first computer system to achieve master-class performance in chess *without* using a supercomputer and without using detailed clues and advice from a human; it *learned* how to play the game at that level.

But all of this is just a beginning. At www.face-rec.org, a series of reviews basically show that two of the three top working systems today rely on neural network concepts (vonderMalsburg and Wechsler). The key to face recognition turns out to be the ability to *learn* new "invariants" or transformations, more complex than simple two-dimensional translation. This offers some easy short-term possibilities: to exploit CSRNs to learn the relevant mappings, which could never be learned before. But it also poses a very fundamental question: how can the *brain* learn such transformations? (The topic of face recognition is also discussed in the chapter by Wong and Cho.)

Here is a new, more formal way to think about what is going on here. The first challenge here is to learn "symmetries of the universe." More concretely, the challenge to the brain is to learn a family of vector maps $f_\alpha$ such that:

$$Pr(f_\alpha(\mathbf{x}(t+1)|f_\alpha(\mathbf{x}(t))) = Pr(\mathbf{x}(t+1)|\mathbf{x}(t))$$ for all $\alpha$ and the same conditional probability distribution Pr.

This new concept may be called stochastic invariance. Intuitively, the idea is that the probability of observing $\mathbf{x}(t+1)$ after observing $\mathbf{x}(t)$ should be the same as the probability of observing the *transformed* version of $\mathbf{x}(t+1)$ after observing the *transformed* version of $\mathbf{x}(t)$. For example, if the probability of observing a red square in the left side of the visual field is 0.25 two seconds after you observe a blue circle in the left side of the visual field, then the probability of observing a red square in the *right* side should also be 0.25 two seconds after you observe a blue circle in the right side, *if* the interchange of left and ride sides is one of the symmetries of nature.

Once a brain learns these symmetries, it may exploit them in one or more of three ways:

(1) "reverberatory generalization": after observing or remembering a pair of data $\{\mathbf{x}(t+1), \mathbf{x}(t)\}$, also train on $\{f_\alpha(\mathbf{x}(t+1)), f_\alpha(\mathbf{x}(t))\}$;
(2) "multiple gating": after inputting $\mathbf{x}(t)$, pick $\alpha$ so as to use $f_\alpha$ to map $\mathbf{x}(t)$ into some canonical form, and learn a universal predictor form canonical forms. (This is analogous to the Olshausen model, which is *very* different in principle from neuroscience models of spontaneous or affective gating and attention.)
(3) "multimodular gating": like multiple gating, except that multiple parallel copies of the canonical mapping are used in parallel to process more than one subimage at a time in a powerful way.

Human brains seem to rely on the first two, or the second. Perhaps higher levels of intelligence could be designed here. But this begs the question: how could these maps be learned? How could the brain learn to map complex fields into a condensed, canonical form for which prediction is much easier to learn? How can the "Objects" in an ObjectNet be learned?

This suggests an immediate and astonishingly simple extension of the ObjectNet theory. In 1992, I proved basic consistency results for a new architecture called the "Stochastic Encoder/ Decoder Predictor" (SEDP).[12, chapter 13]. SEDP *directly* learns condensed mappings. It is an adaptive nonlinear generalization of Kalman filtering, explicit enough to allow the learning of symmetry relations. As with the earlier HDP and CSRN architectures, it will require many specific tricks to improve its learning speed. (e.g., exploitation of nearest neighbor relation in the learning, and salience flows?). It provides a principled way to *learn* the symmetry groups which are the foundation for a principled approach to spatial complexity.

# APPENDIX: MORE LINKS TO THE BRAIN AND TO HUMAN SUBJECTIVE EXPERIENCE

*A. The Larger Context*

The text of this chapter *mainly* tries to look down on the most basic mammal brain – "the soulless rat" – and understanding it in engineering terms. It tries to unify an optimization approach – and the engineering needed to work out the mathematical and function details – with cognitive or systems neuroscience. In a way, it is providing a pathway for trying to truly unify computational neuroscience with systems and cognitive neuroscience. That specific effort at unification is a great, concrete opportunity for a major new stream of scientific research. My comments about subjective human experience in that main text were really just heuristics to aid understanding. It is important to keep straight what is or is not part of that new stream of science.

At the same time, we as scientists or as humans have a larger agenda. In this appendix, I will talk more about that larger agenda. As humans, we can look at ourselves in the mirror, objectively, using the same scientific tools we would use to look at a mouse or a rat. But we can also look at our own inner experience. The larger challenge here is to arrive at an understanding which can fit *three* sources of empirical data or validation – the data from ordinary experiments on brain and behavior, seen objectively; the data from subjective experience, seen within ourselves; and the data we get from systematically testing the *functional* capabilities of our models/designs when applied to complex tasks, as in engineering. This kind of unified understanding is what we really need in order to better understand ourselves. The chapters by Kozma, Levine and Perlovsky address the higher cognitive functions of the human mind.

But how can we arrive at such a unified understanding, when one side is struggling hard to rise as high as a mouse, and the other is well beyond it?

In parallel to the scientific track, which this chapter has focused on, the new theory opens up a new humanistic track, which is less quantitative for now, but broader in scope. The human mind includes everything I just talked about regarding the "mouse brain," but includes two major extensions or refinements, in my view: (1) a partially evolved new capacity for symbolic reasoning, language and empathy, based in part on "mirror neurons" and new structures in the NRTP (a small but important sheath around the thalamus); and (2) a kind of embryonic collective intelligence effect, which goes far beyond what we easily see in the laboratory. This humanistic track has much to gain from the scientific track, because *everything we see in the mouse brain is also present in us* (even if we do sniff less often).

Again, I will only give a few new highlights here. I have posted a number of papers on various aspects of the scientific and humanistic track at www.werbos.com. Several of those papers include flowcharts associating components of the brain with components of an integrated ADP design. The chapter by Levine also discusses relations between cognitive models and modules in the brain.

### B. *Comparative Cognitive Neuroscience*

At the start of this chapter, I briefly referred to the classic, important work of M.E. Bitterman, who demonstrated major *qualitative* differences in intelligence between the reptile, the bird, the mammal, and so on. The "Rockefeller" series on the brain, *The Neurosciences* edited by Schmitt and Worden, also contains a beautiful summary of the differences in the "wiring diagrams" between these different levels of brain.

The key unique feature of the mammal brain is the *six-layer* type of cortex, the "neocortex," which accounts for more than half the weight of a human brain. This new structure basically evolved as a *merger* of two different three-layer cortices from lower organisms. This has important connections to the story of this chapter.

Even the bird brain has an ability to handle spatial complexity far beyond what our engineering systems can do today. After all, complex tasks in image processing and spatial "reasoning" are essential to the life of a bird. Thus I would claim that the Stochastic Encoder/Decoder Predictor (SEDP) structure I mentioned in Section IV exists even within the three-layer cortex of the bird which performs such tasks.

Like Kalman filtering, SEDP tries to estimate or impute a representation of the true state of the external variable, by estimating (or "filtering") its value $R_i$ in real time. For each such estimated variable, SEDP has to consider three different numerical values at each time – a value *predicted* from the past, a value *estimated* by accounting for new sensory data, and a value *simulated* in order to reflect the uncertainty of the estimate. How could a brain structure maintain a coordination of three different numerical values, which fits the required calculations? The obvious way would be to evolve a very large kind of neuron which first computes the prediction in its first section, and then tests the prediction and updates it in a connected large second section. And then adds some random noise at a final, third stage. Strange as this may sound, it fits beautifully with the actual structure of giant pyramid neurons, which are found even in the three-layer cortex of the bird. This also requires a kind of clock pulse to synchronize the calculations, controlling the interface where the prediction enters the estimation part of the neuron; in fact, Scheibel and Scheibel long ago reported such a nonspecific modulation input touching the middle part of the giant pyramid cells in the human six-layer cortex.

This chapter has argued that we can build ADP systems which are very effective in coping with spatial complexity, even before we do all we can with imagination/creativity and temporal complexity. Could it be that nature has already done the same thing, with the bird or even with vertebrates lower than the bird?

The figure on creativity in section IV illustrates a further idea from [21]. On that slide, you can see an arrow going from Layer V of the neocortex down to the basal ganglia. The idea here is as follows. When the two early three-layer cortices became fused, in the mammal, the *stochastic* aspect of the SEDP design became harnessed to solve an additional task: the task of stochastic search or creativity in making decisions. In effect, layer V of the mammal brain provides a new capability to *suggest options* for decisions. Is it a case where Layer V proposes and the basal ganglia disposes? Perhaps. And perhaps a powerful imagination or creativity is what mammal brains have that lower brains do not. (Lower vertebrates may of course possess simpler, less powerful stochastic exploration mechanisms, similar to the earlier work by Barto, Sutton and Anderson on reinforcement learning, or to the stochastic stabilization mechanisms discussed in the later sections of [16].)

But before that kind of creativity or imagination can evolve, there must first be some wiring to implement the concept of "decision." That is where complexity in time comes into it. Complexity in time is the main focus of [21] and [22]. Some tutorial flow charts and slides about this are also posted, in the later sections of the tutorials posted on my web page. Here again, I will only present some new thoughts.

At the IJCNN05 conference in Montreal, the neuroscientist Petrides presented some important and persuasive new results on the "advanced" parts of the human neocortex. Here is a simplified version of his conclusions: "I have thoroughly reviewed and re-assessed the experiments of my illustrious predecessors, like Goldman-Rakic, and discovered that their interpretations were not quite right. Today, I will speak about the two highest parts of the human cerebral cortex, the dorsolateral prefrontal cortex and the orbitofrontal. It turns out that these two regions are there to allow us to answer the two greatest challenges to the human mind: 'Where did I leave my car this time in the parking lot? And what was I trying to do anyway?' "

More concretely – these parts of the cerebral cortex of the human connect directly to the basal ganglia, where decisions are "engaged." Based on classic artificial intelligence, some neuroscientists once speculated that the brain contains a kind of explicit hierarchy of decisions within decisions, used to structure time into larger and larger intervals. They speculated that different connections to different parts of the basal ganglia might refer to different levels of the temporal hierarchy. But recent studies of the basal ganglia do not fit that idea. Instead, they suggest that the "hierarchy in time" is more implicit and deliberately fuzzy. (I received incredible hate mail from people claiming to represent the American Association for Artificial Intelligence when Lotfi Zadeh and I suggested a need for a mathematically well-defined fuzzy relationship here!) Petrides' work suggests that the connection from dorsolateral cortex to the basal ganglia proposes the "verb" or "choice of discrete decision block type" to the basal ganglia. It also suggests that the "hierarchy" is implicit and fuzzy, based on how one decision may engage others – but may in fact be forgotten at times because of limited computational resources in the brain. (Indeed, in the real world of human life, I have seen many, many examples of people producing poor results, because they seemed to forget what they were really trying to do, or why. People do crash cars and they do crash entire nations.) Perhaps, in the human, the orbitofrontal or another region tends to propose the "object" or parameters of the decision to be made. Yet I would guess that even birds can make decisions (albeit unimaginative decisions), and possess similar connections to their own well-developed basal ganglia from more primitive cerebral cortex.

A discussion of brain evolution would not be complete without some reference to the great evolutionary biologist George Gaylord Simpson. Simpson pointed out that we should not assume that all the attributes of modern mammals evolved at the same time. Looking at the indentations of early mammal skulls, he postulated that the modern mammal brain evolved much later than the other attributes. The evolutionary niche of the fast breeding, teeming mice living under the dinosaurs allowed relatively rapid evolution – but it still required many millions of years, because of the complexity of the wiring that had to be changed. Could it be that the breakthrough in the creativity and imagination of the mouse was what forced the later dinosaurs (like the Maiasaurus) to put more energy into protecting their eggs – and weakened them enough to allow their easy demise when confronted with environmental stress? Humans today, like that early mouse, also represent the early stages of the kind of "quantum break-through" that Simpson discussed. Alternatively – as I submit this paper, it has been discovered that the complex mammals which radiated out from a small mouse-like creature, after the dieoff of the dinosaurs, were *themselves* replaced by a new family of mammals, radiating out from another small mammal. Could it be that evolution of the brain was the crucial missing factor here?

*C. Links to Subjective Experience*

Scientists need to know how to give up empathy, on a temporary and tentative basis, in order to map out and improve what they can learn from laboratory experiments in neuroscience and psychology and engineering as clearly as possible. But if human decision-makers did that all the time, they would be well on track towards blowing up the world. Furthermore, by unifying our subjective experience and our scientific understanding, we can arrive at important new testable insights in all of these realms. When we totally give up empathy, we are basically renouncing the use of our own mirror neurons, and lowering the level of intelligence that we express in our work.

The well-known neuroscientist and psychologist Karl Pribram has often made this point in more graphic terms. "Animal behavior is really all about the four f's – food, fear, … Any so-called model of human intelligence that doesn't give a central place to them is totally crazy." A "theory of intelligence" that does not address emotions is actually more useful as a specimen of human immaturity and neurosis than as a scientific theory. Isaac Newton – a true scientist – was willing to overcome his squeamishness, and insert a needle into his own eye and observe what he saw, in order to test his new ideas about color and light. To better understand the mind, we need to overcome our own squeamishness and look more forcefully into our own subjective experience. The chapters by Levine and Perlovsky also discuss the central role of emotions as a part of human intelligence.

To understand the full database of human subjective experience, we ultimately need to consider, especially, the work of Sigmund Freud, Carl Jung and the larger streams of "folk psychology" and diverse civilizations found all across the streams of human culture. My web page does make those connections, though of course they are not so concise and compartmentalized as an engineering design can be. The chapter by Perlovsky discusses Jung's concepts of differentiation and synthesis as mechanisms of cultural volition.

The work of Freud is especially important to the theory/model presented here, because it was one of the sources which inspired it in the first place. In my earliest paper on the idea of ADP[4], I proposed that we approximate the Bellman equation somehow to develop a working mathematical design/theory – but I did not know how. I also suggested that Freud's theory of "psychodynamics" might turn out to be right, and to work, if only we could translate it into mathematics. The workable 1971-1972 design shown above – the first true adaptive critic design – was the result of my going ahead and doing what I proposed in that paper. (I hope that some of you, dear reader, will be able to do the same for what I propose here, since I do not have as many years left to me, and have other life-or-death responsibilities as well.) Instead of letting myself be overcome by squeamishness at Freud's fuzzy language, I looked hard and figured out how to translate it into mathematics. That is what backpropagation really is – a translation into mathematics (and generalization) of Freud's concepts of psychodynamics.

Freud had many other insights which are important to the program of research proposed here. For example, when he discusses the "struggle" between the "ego" and the "id," what is he really talking about? In that discussion, his "id" is really just the part of the brain that uses associative memory or nearest-neighbor associations to make predictions. (For example, if someone wearing a blue shirt hits you, when you are a child, you may feel full of fear years later whenever you see a person wearing a blue shirt. That's an id-iotic response.) The "ego" includes a more coherent kind of global cause-and-effect style of prediction, like what we get in engineering when we train Time-Lagged Recurrent Network using the kind of tools that Ford has used, combining back-propagation and DEKF methods. Even when we try to build a "second generation brain" – a brain far simpler than the brain of a bird, let alone a mouse – it is essential that we learn how to combine these "id" and "ego" capabilities together. This is the basis for my proposal for "syncretism," discussed in various papers in my web page. The work to implement and understand syncretism is part of the research described in section III – even before we reach so high as the brain of the bird.

Neural network designs used in engineering today are very different from most of the models in computational neuroscience (CNS). The reason is that most people in the CNS niche today rely heavily on Hebbian models, which is really powerful only for creating memory. Some CNS people have suggested that the entire brain is nothing but a memory machine – a suggestion just as absurd as the AI idea of a brain without emotions; it is another example of a theory motivated by a lack of powerful enough mathematical tools to cope with the basic empirical reality which stares us in the face. Engineering today mainly focuses on the larger-scale systems tools which address the "ego" aspect. To build the best possible "second generation" brain, we need to unify these two aspects, exactly as Freud and I have proposed. Those aspects also exist in the highest levels of intelligence in the human mind and human social existence. The chapters

D. *A Humorous Postscript on an Engineering Application*

As I finish this chapter, I can imagine someone asking: "Dr. Werbos, are you really trying to tell us that some kind of automated bird brain really could improve on what we are doing today to control the East Coast electric power grid? That is a huge insult to the thousands of brilliant engineers, economists, and lawyers who have devised that system."

Yes, I am saying that a kind of automated bird brain could do better than a lot of what we are doing today, but it is not intended as an insult. Here are some reasons why it is possible.

First, in controlling the power grid, we rely heavily on *automated* systems, in part because events happen too fast for humans to keep up with at the lowest levels. It also helps to be able to communicate at the speed of light. A distributed, automated bird brain would be far more advanced than what we have today at the high-bandwidth aspect of control – if it is properly interfaced with the other parts.

Second, the best system we have today to manage most aspects of the grid is called Optimal Power Flow (OPF). By optimizing the entire system as one large system, it provides integrated results much better than we could get from separately managed independent components all working on their own. OPF is the most widely used university-developed algorithm in use in electric utilities today. But today's OPF lacks the ability to *anticipate* future possible developments. Electric power grids have large and immediate needs to inject more foresight into the automated systems, for many reasons. (e.g. anticipating and avoiding possible "traffic jams," efficient "time of day pricing," shifting sources and loads between day and night, especially as solar power and plug-in hybrid cars become more popular.) Traditional second-generation ADP can provide the necessary foresight, but cannot handle all the complexity. This is why spatial complexity – including ObjectNets – is an essential think to add to ADP. Of course, the nodes in such a distributed network might well need to use new chips, like the Cellular Neural Network chips now being manufactured for use in image processing, in order to make fullest use of the new mathematical designs.

This extension of OPF – called "Dynamic Stochastic OPF" – is discussed at length in chapters by myself and by James Momoh in [7]. (Momoh developed the original version of OPF distributed to electric utilities by the Electric Power Research Institute, which cosponsored the workshop on DSOPF which led to those chapters.)

*E. The underlying mechanisms of creativity and imagination*

In mathematical terms, creativity and imagination are linked to *nonconvex* optimization, the ability of a system to jump from one strategy of action or situation to a better situation, even when there is no gradual incremental path from the old to the new. In simplified accounts of learning, I usually say that a complete intelligent system needs two major components: (1) an *incremental* learning system, which must use derivatives and gradients in some form to point the way "up" to better outcomes; and (2) a creativity system, which is not always guaranteed to work and must include stochastic elements, to try to find higher mountains beyond the hill where one presently exists. The key challenge with creativity is to figure out how organisms can learn, incrementally, to be more effective in creativity, in exploring the range of possible options.

Sometimes, I go on to say that nature must provide us with multiple creativity systems, because creativity is so difficult and imperfect, requiring multiple levels of creativity. For humans, the most important sources of creativity involve the use of symbolic reasoning (mainly words and mathematics) and various collective intelligence kinds of mechanisms. But for the individual mouse brain (which is already far more creative than the lizard), I associate the main creativity mechanism with layer V of the neocortex, as in the figure above. Layer V is where stochastic effects are most obvious, in driving outputs of the neocortex which go directly to the basal ganglia, where decisions are engaged into commitment and action.

A key aspect of this mechanism is that it is unique to mammal brains. Reptile brains do not have it – but they have a kind of precursor to it [2]. The most important challenge to comparative cognitive neuroscience today is to understand exactly what happened when that precursor evolved into the modern mouse brain. George Gaylord Simpson found evidence long ago that the evolution of the neocortex occurred long after the initial evolution of the earliest warm-blooded proto-mammals. He suggested that this event may have had larger implications than we appreciate even today. Perhaps the creativity of mice and rats in learning how to eat dinosaur eggs was one of the important factors in changing the general trends of the time.

But what does this mechanism really do, in cybernetic or engineering terms?

The ADP kinds of designs which I have discussed do not have the same degree of local minimum problems that more conventional neural network algorithms do. For example, if you are standing on top of a hill, looking at a distant mountain… and if your goal is to be as high as possible (i.e. your utility function U = your height), the Bellman equation points you towards a path up the mountain, even if you must first go downhill. More precisely, no path from your present location to the top of the mountain is monotonic in

U. However, there *is* a path which is monotonic in J. If you learn a good enough approximation to the J function, and you maximize J in the near-term future, then you will never be stuck in a local minimum. In theory, this is a perfect solution to the local minimum problem.

As a practical matter, however, there are two major difficulties in implementing this kind of theory. These, in turn, lead to a requirement for two kinds of new components in a learning system. Please forgive me if I name these components (1) "imagination"; and (2) "creativity proper."

"Imagination" is the neural circuitry which generates images or frames of possible states of reality (**R**) used, above all, for improving the training of the Critic networks of the brain. Our estimated J functions cannot really reflect the opportunity of the mountain in the distance, unless we train them over points in the region which include the mountain and enough of what connects us to the mountain. In my 1987 SMC paper [17], I did stress the importance of this kind of "dreaming" or simulation in training an ADP system. I also discussed it at great length with Richard Sutton in 1988; in 1990, he included a chapter on "dyna" simulations embodying that principle in our joint book.[18] However, I did not discuss the details of how the brain could learn to be ever more effective over time in dreaming. LaBerge of Stanford, one of the world's leading sleep researchers [25], has also proposed that the main function of dreams is *simulation*.

"Creativity proper" is the circuitry which addresses the challenge of nonconvex optimization which remains even after one has properly trained the Critic networks. In first generation brains, it addresses the maximization problem of finding the vector **u**(t) which really maximizes the expected value <U(t) + J(t+1)>. Usually, nonconvexity is not such a big problem at that level, because the gradient of J tells us where we should be going, and we can simply turn (monotonically) in the right direction, when we know what it is. However, *after we reach the stage of evolution* when we use "decision blocks" to cope with temporal complexity, we face the new challenge of what *decision* to make. *Decisions* are not really the same as action vectors **u**. For example, when we play a game of Go or of chess, we *decide* at each turn which piece to move, where. That is not the same as deciding how hard to push our wrist muscles at the present instant. It is also a highly nonconvex problem. "Imagination" is part of the judgmental Critic faculty, which helps us evaluate and compare possible moves. "Creativity proper" helps us see what the interesting moves are to evaluate. The two faculties are closely linked in many ways, but they are distinct.

In [26,21], I have discussed this challenge in more detail, with some crude place-holder models for "Option Network" adaptive circuits. Just in the past few years, new research has begun to appear which addresses this challenge more systematically; for example, Pelikan and Venagamoorthy have begun to explore how the learning of probability distributions (pdf) can guide and improve nonconvex search. In the context of electric power, Venagamoorthy has said that the learning of pdf does significantly improve the results of Particle Swarm Optimization, which is currently the most effective method in those applications (where the decisions revolve about continuous variables, not the binary variables which conventional genetic algorithms address).

More generally – if we think about how we, as humans, find our way to "distant mountains" in the space of possibilities, an additional thought comes to mind. As early as the bird (and perhaps earlier, in the ladder of vertebrate brains), we evolved a complex mechanism for developing "spatial maps," closely related to the Object Net designs mentioned above. Before the evolution of the mammal brain, the reptile brain contained two *three-level* cortices, one of which assisted in coping with spatial complexity. Could it be that mammal brains *built on* that mechanism even more than I described above? Could it be that the mechanism which was formerly used to explore and map physical space was then linked up in a new way, harnessing it to exploring the space of possibilities – a kind of cognitive map of "state space"? Is this the key to the superior creativity and imagination of mammals over reptiles? How does this connect to what has been seen in EEG data on sleep states in reptiles in birds – if such exists? These are important questions we should be working to address as soon as possible in our research – but they start out from assuming a foundation in spatial complexity, which is perhaps the most urgent priority for research here.